\documentclass[12pt]{article}

\usepackage{amssymb}
\usepackage{amsmath}
\usepackage{amsfonts}

\newcommand{\C}{\mathbb{C}}

\newcommand{\be}{\begin{equation}}
\newcommand{\bea}{\begin{eqnarray}}
\newcommand{\eea}{\end{eqnarray}}

\newcommand{\kt}{\rangle}
\newcommand{\br}{\langle}

\newcommand{\ed}{\end{document}}

\begin{document}

\title{Classical and Quantum Fermions Linked by an Algebraic
Deformation}
\author{Ali Mostafazadeh\thanks{E-mail address:
amostafazadeh@ku.edu.tr}\\ \\
Department of Mathematics, Ko\c{c} University,\\
Rumelifeneri Yolu, 80910 Sariyer, Istanbul, Turkey}
\date{ }
\maketitle

\begin{abstract}
We study the regular representation $\rho_\zeta$ of the
single-fermion algebra ${\cal A}_\zeta$, i.e., $c^2=c^{+2}=0$,
$cc^++c^+c=\zeta~1$, for $\zeta\in [0,1]$. We show that $\rho_0$
is a four-dimensional nonunitary representation of ${\cal A}_0$
which is faithfully irreducible (it does not admit a proper
faithful subrepresentation). Moreover, $\rho_0$ is the minimal
faithfully irreducible representation of ${\cal A}_0$ in the sense
that every faithful representation of ${\cal A}_0$ has a
subrepresentation that is equivalent to $\rho_0$. We therefore
identify a classical fermion with $\rho_0$ and view its
quantization as the deformation: $\zeta:0\to 1$ of $\rho_\zeta$.
The latter has the effect of mapping $\rho_0$ into the
four-dimensional, unitary, (faithfully) reducible representation
$\rho_1$ of ${\cal A}_1$ that is precisely the representation
associated with a Dirac fermion.
\end{abstract}
%\vspace{2mm}
%PACS numbers: 03.65.Bz\\
%\vspace{2mm}

%\baselineskip=24pt

\section{Introduction}

The description of fermions in terms of the Clifford algebra
relations
    \bea
    &&cc^++c^+c=1,
    \label{xc1}\\
    &&c^2=c^{+2}=0,
    \label{xc2}
    \eea
dates back to early days of quantum physics. This algebra may be
obtained by quantizing a classical system with fermionic
variables, e.g., a free fermion or a fermionic oscillator
\cite{berezin,dewitt}. The classical fermionic variables  satisfy
the Grassmann algebra relations
    \bea
    &&cc^++c^+c=0,
    \label{g1}\\
    &&c^2=c^{+2}=0.
    \label{g2}
    \eea
Therefore similarly to the case of bosonic variables, the
quantization of a fermionic variable may be viewed as the
deformation of the algebraic relations
    \bea
    &&cc^++c^+c=\zeta 1,
    \label{c1}\\
    &&c^2=c^{+2}=0,
    \label{c2}
    \eea
where the deformation parameter $\zeta$ takes values in $[0,1]$.
Motivated by the method used in \cite{jpa01c} to study the
representation theory of orthofermions, we investigate in this
paper the effect of the deformation $\zeta\to 0$ on the
representations of the associative algebra ${\cal A}_\zeta$
generated by $1,~c$, and $c^+$ and subject to relations (\ref{c1})
and  (\ref{c2}).

It is well-known \cite{spin-geometry,jpa01c} that the
representations of the Clifford algebra ${\cal A}_1$ are, up to
equivalence, direct sums of copies of the trivial representation
$\rho_{\rm trivial}$:
    \[\rho_{\rm trivial}(1)=\rho_{\rm trivial}(c^+)=
    \rho_{\rm trivial}(c)=0,\]
and the two-dimensional unitary (or $*$-) representation
$\rho_\star$:
    \[\rho_\star( 1)=\left(\begin{array}{cc}
        1&0\\
        0&1\end{array}\right),~~~~
           \rho_\star(c)=\left(\begin{array}{cc}
        0&1\\
        0&0\end{array}\right),~~~~
           \rho_\star(c^+)=\left(\begin{array}{cc}
        0&0\\
        1&0\end{array}\right)= \rho_\star(c)^\dagger.\]
For $\zeta\neq 0$, one can simply absorb the deformation parameter
$\zeta$ in the definition of $c$ and/or $c^+$. Therefore the
representations of ${\cal A}_\zeta$ for $\zeta\neq 0$ are the same
as those of ${\cal A}_1$. As we shall see below, for $\zeta=0$ the
situation is completely different.

Before, we begin our analysis, we wish to make note of the
following facts about the Grassmann algebra ${\cal A}_0$.
    \begin{enumerate}
    \item ${\cal A}_0$ does not admit nontrivial unitary
    representations. In order to see this we first note that in view
    of Eqs.~(\ref{c1}) and  (\ref{c2}) the algebra ${\cal A}_\zeta$
    is spanned by the basis elements $1,c^+,c$ and $n$, where
    $n:=c^+c$. Now, let $({\cal H},\br~,~\kt)$ be an inner-product
    space and $\rho:{\cal A}_0\to {\rm End}({\cal H})$ be a
    representation of ${\cal A}_0$ where `End' abbreviates
    `Endomorphism' (a linear operator mapping ${\cal H}$ into
    ${\cal H}$.) By definition, if $\rho$ is a unitary
    representation, then $\rho(c^+)=\rho(c)^\dagger$, where a
    dagger stands for the adjoint of the corresponding operator.
    According to Eqs.~(\ref{g1}) and (\ref{g2}), the unitarity of
    $\rho$ implies for all $|\psi\kt\in{\cal H}$,
        \[||\rho(n)|\psi\kt||^2=\br\psi|\rho(n)^\dagger\rho(n)
        |\psi\kt=\br\psi|\rho(n)^2|\psi\kt
        =\br\psi|\rho(n^2)|\psi\kt=0.\]
    Hence $\rho(n)|\psi\kt=0$. On the other hand,
        \[||\rho(c)|\psi\kt||^2=\br\psi|\rho(c)^\dagger\rho(c)
        |\psi\kt=\br\psi|\rho(n)|\psi\kt=0.\]
    Therefore for all $|\psi\kt\in{\cal H}$, $\rho(c)|\psi\kt=0$, so
    that $\rho(c)=0$, $\rho(c^+)=0$, and $\rho$ is trivial.
    \item
The only irreducible representation of ${\cal A}_0$ is the
one-dimensional representation defined by
    \be
    \rho^{(1)}_\emptyset(1)=1,~~~~\rho^{(1)}_\emptyset(c^+)=\rho^{(1)}_\emptyset(c)=0.
    \label{rho-1}
    \end{equation}
To see this let $\rho:{\cal A}_0\to {\rm End}(V)$ be an arbitrary
representation. Then $V_\emptyset={\rm
Im}(\rho(n)):=\{\rho(n)v|v\in V\}$ is an invariant ($\rho$-stable)
subspace \cite{fell-doran}, because for all $x\in {\cal A}_0$ and
for all $v\in V_\emptyset$, $\rho(x)v\in V_\emptyset$. This shows
that $\rho$ is reducible. Furthermore, the subrepresentation
obtained by restricting $\rho$ to $V_\emptyset$ is clearly
equivalent to $\rho^{(1)}_\emptyset$.
    \end{enumerate}

Next, consider the regular representation $\rho_\zeta:{\cal
A}_\zeta\to{\rm End}({\cal A}_\zeta)$ of ${\cal A}_\zeta$ that is
defined by
    \be
    \forall x,y\in {\cal A}_\zeta,~~~~~\rho_\zeta(x)y:=xy.
    \label{can}
    \end{equation}
In the basis $\{1,c^+,c,n\}$, where
    \[1=\left(\begin{array}{c}
        1\\0\\0\\0\end{array}\right),~~~
    c^+\left(\begin{array}{c}
        0\\1\\0\\0\end{array}\right),~~~
    c=\left(\begin{array}{c}
        0\\0\\1\\0\end{array}\right),~~~
    n=\left(\begin{array}{c}
        0\\0\\0\\1\end{array}\right),\]
we have
    \bea
    &&\rho_\zeta(1)=\left(\begin{array}{cccc}
        1&0&0&0 \\
        0&1&0&0\\
        0&0&1&0\\
        0&0&0&1\end{array}\right),~~~~
    \rho_\zeta(c^+)=\left(\begin{array}{cccc}
        0&0&0&0 \\
        1&0&0&0\\
        0&0&0&0\\
        0&0&1&0\end{array}\right),
    \label{rep1}\\
    &&\rho_\zeta(c)=\left(\begin{array}{cccc}
        0&\zeta&0&0 \\
        0&0&0&0\\
        1&0&0&\zeta\\
        0&-1&0&0\end{array}\right),~~~~
    \rho_\zeta(n)=\left(\begin{array}{cccc}
        0&0&0&0 \\
        0&\zeta&0&0\\
        0&0&0&0\\
        1&0&0&\zeta\end{array}\right).
    \label{rep2}
    \eea
Here we have made use of Eqs.~(\ref{c1}), (\ref{c2}), and
(\ref{can}).

It is not difficult to show that $\zeta\neq 0$ if and only if
$\rho_\zeta$ is a pseudo-unitary representation \cite{p8}. This is
equivalent to the requirement that there is a linear Hermitian
invertible operator $\eta$ such that
    \be
    \rho_\zeta(c^+)=\rho_\zeta(c)^\sharp:=
    \eta^{-1}\rho_\zeta(c)^\dagger\eta.
    \label{pseudo}
    \end{equation}
This can be easily checked by taking $\eta$ to be an arbitrary
$4\times 4$ matrix and imposing the condition
$\eta\rho_\zeta(c^+)=\rho_\zeta(c)^\dagger\eta$ to determine the
matrix elements of $\eta$. It follows that the determinant of
$\eta$ is proportional to $\zeta$. Therefore $\rho_0$ is not
pseudo-unitary. For $\zeta\neq 0$ there are many invertible
matrices $\eta$ satisfying (\ref{pseudo}), e.g.,
    \be
    \eta=\left(\begin{array}{cccc}
        0&\zeta^{-1}&\zeta^{-1}&0   \\
        \zeta^{-1}&0&0&1\\
        \zeta^{-1}&0&0&0\\
        0&1&0&0\end{array}\right).
    \label{eta}
    \end{equation}
Furthermore, in this case, there are similarity transformations
    \be
    \rho_\zeta(x)\to \rho'_\zeta(x):=S^{-1}\rho_\zeta(x)S
    \label{sim}
    \end{equation}
that reduce $\rho_\zeta$ into the direct sum of two nontrivial
two-dimensional irreducible representations. A convenient choice
is
    \be
    S=\left(\begin{array}{cccc}
        \zeta&0&0&0 \\
        0&1&0&0\\
        0&0&1&0\\
        -1&0&0&1\end{array}\right).
    \label{s}
    \end{equation}
Using Eqs.~(\ref{rep1}), (\ref{rep2}), (\ref{sim}), and (\ref{s}),
we have
    \bea
    &&\rho_\zeta'(1)=\left(\begin{array}{cccc}
        1&0&&   \\
        0&1&& \\
        &&1&0\\
        &&0&1\end{array}\right),~~~~
    \rho_\zeta'(c^+)=\left(\begin{array}{cccc}
        0&0&&   \\
        \zeta&0&&\\
        &&0&0\\
        &&1&0\end{array}\right),
    \label{rep3}\\
    &&\rho'_\zeta(c)=\left(\begin{array}{cccc}
        0&1&&\\
        0&0&&\\
        &&0&\zeta\\
        &&0&0\end{array}\right),~~~~
    \rho'_\zeta(n)=\left(\begin{array}{cccc}
        0&0&&\\
        0&\zeta&&\\
        &&0&0\\
        &&0&\zeta\end{array}\right),
    \label{rep4}
    \eea
where the empty entries are zero. Clearly $\rho'_1$ is the direct
product of two copies of the basic unitary representation
$\rho_\star$ of the Clifford algebra ${\cal A}_1$. Also note that
for $\zeta=0$ the matrix $S$ is not invertible, and the above
construction does not apply.

In fact, it is not difficult to show that the Grassmann algebra
${\cal A}_0$ does not admit one, two, or three-dimensional
representations that are faithful. In order to see this, consider
an arbitrary representation $\rho:{\cal A}_0\to{\rm End}(V)$ where
$V$ is a complex (or real) vector space, and suppose that $\rho$
is faithful (one-to-one). Then there is $v_1\in V$ such that
$v_4:=\rho(n)v_1\neq 0$. This together with the fact that
$\rho(n)=\rho(c^+)\rho(c)$ imply $v_2:=\rho(c^+)v_1\neq 0$ and
$v_3:=\rho(c^+)v_1\neq 0$. Next let $\lambda_i\in\C$, with
$i\in\{1,2,3,4\}$, satisfy
    \be
    \sum_{i=1}^4 \lambda_i\,v_i=0.
    \label{li}
    \end{equation}
Applying $\rho(n)$ to both sides of this equation yields
$\lambda_1=0$. Substituting this equation in (\ref{li}) and acting
by $\rho(c)$ and $\rho(c^+)$ on both sides of the resulting
equation lead to $\lambda_2=0$ and $\lambda_3=0$, respectively.
Therefore $\lambda_i=0$ for all $i\in\{1,2,3,4\}$; $v_i$ are
linearly independent, and dim$(V)\geq 4$. This in particular shows
that the regular representation $\rho_0$ is the `lowest'
dimensional faithful representation. In the following we shall use
the term `{\em faithfully irreducible representation}' by which we
mean a faithful representation that does not admit a proper
faithful subrepresentaion. Note that a faithfully irreducible
representation may very well be reducible. The typical example is
the regular representation $\rho_0$.

Next, consider the span of $v_i$:
    \[V_{v_1}:={\rm Span}(v_1,v_2,v_3,v_4)=\left\{\left.
    \sum_{i=1}^4   \lambda_i\,v_i~\right|~\lambda_i\in\C\right\}.\]
It is not difficult to see that for all $v\in V_{v_1}$ and $x\in{\cal A}_0$, $\rho(x)v\in V_{v_1}$. Hence the restriction $\rho_{v_1}:{\cal A}_0\to{\rm End}(V_{v_1})$ of $\rho$ to $V_{v_1}$, which is defined by
    \[\forall x\in{\cal A}_0~~{\rm and}~~\forall v\in V_{v_1},
    ~~~~\rho_{v_1}(x)v:=\rho(x)v,\]
provides a representation of ${\cal A}_0$. Clearly, $\rho_{v_1}$
is equivalent to the regular representation $\rho_0$. This proves
the following.
    \begin{itemize}
    \item[~] {\bf Theorem:} {\em Every faithful representation of
    the Grassmann algebra ${\cal A}_0$ has a subrepresentation that
    is equivalent to the regular representation $\rho_0$. In
    particular, $\rho_0$ is (up to equivalence) the unique
    4-dimensional faithfully irreducible representation of
    ${\cal A}_0$.}
    \end{itemize}
This is analogous to the well-known fact about the Clifford
algebra ${\cal A}_1$, namely that every faithful representation of
${\cal A}_1$ has a subrepresentation that is equivalent to the
canonical representation $\rho_\star$. In particular, $\rho_\star$
is (up to equivalence) the unique 2-dimensional faithful
irreducible representation of ${\cal A}_1$. However there is a
stronger result \cite{jpa01c} indicating that every representation
of ${\cal A}_1$ is a direct product of copies of the trivial
representation $\rho_{\rm trivial}$ and the canonical
representation $\rho_\star$. A similar result does not hold for
${\cal A}_0$. This is mainly because there are, besides the
trivial representation, one, two and three-dimensional nonfaithful
representations, namely $\rho_\emptyset^{(1)}:{\cal A}_0\to {\rm
End}(\C)=\C$ of (\ref{rho-1}) and $\rho_\emptyset^{(2)}:{\cal
A}_0\to {\rm End}(\C^2)$ and $\rho_\emptyset^{(3)}:{\cal A}_0\to
{\rm End}(\C^3)$ defined by
    \bea
    \rho_\emptyset^{(2)}(1)=1,~~~~\rho_\emptyset^{(2)}(c)=0,
    ~~~~\rho_\emptyset^{(2)}(c^+)=\mu,
    \label{rho-2}\\
    \rho_\emptyset^{(3)}(1)=1,~~~~\rho_\emptyset^{(3)}(c)=\nu,
    ~~~~\rho_\emptyset^{(3)}(c^+)=\nu^+,
    \label{rho-3}
    \eea
where
    \be
    \mu:=\left(\begin{array}{cc}
    0&1\\
    0&0\end{array}\right),~~~~~
    \nu:=\left(\begin{array}{ccc}
    0&1&0\\
    0&0&0\\
    0&0&0\end{array}\right),~~~~~
    \nu^+:=
    \left(\begin{array}{ccc}
    0&0&1\\
    0&0&0\\
    0&0&0\end{array}\right).
    \end{equation}

In view of the above-stated uniqueness property of the regular
representation $\rho_0$ of the Grassmann algebra ${\cal A}_0$, we
propose to identify a `classical fermion' with $\rho_0$. Then the
quantization of $\rho_0$ may be viewed as the deformation
$\zeta:0\to 1$ of the regular representation $\rho_\zeta$ of the
one-fermion algebra ${\cal A}_\zeta$ that maps the classical
fermion $\rho_0$ to the `quantum fermion' $\rho_1$. The latter is
a four-dimensional unitary reducible representation of the
Clifford algebra ${\cal A}_1$ that is associated with a Dirac
fermion. In this sense Dirac fermions are naturally linked with
the quantization of the classical fermions.

$$
***************************
$$

This work has been supported by the Turkish Academy of Sciences in
the framework of the Young Researcher Award Program
(EA-T$\ddot{\rm U}$BA-GEB$\dot{\rm I}$P/2001-1-1).

\ed